\documentstyle[12pt]{article}
\global\parskip 6pt

\normalsize

\def\br{\begin{thebibliography}{abc}}
\def\er{\end{thebibliography}}

\def\be{\begin{equation}}
\def\ee{\end{equation}}
\def\bea{\begin{eqnarray}}
\def\eea{\end{eqnarray}}

\begin{document}
\begin{flushright}
\hfill{SINP/TNP/2010/02}\\
\end{flushright}
\vspace*{1cm}
\thispagestyle{empty}
\centerline{\bf The Central Charge of the Warped $AdS^3$ Black Hole}
\smallskip
\begin{center}
Kumar S. Gupta$^a$\footnote{Email: kumars.gupta@saha.ac.in}, E. Harikumar$^b$\footnote{Email: harisp@uohyd.ernet.in}, Siddharhta Sen$^{c,d}$\footnote{Email: sen@maths.ucd.ie} and M. Sivakumar$^b$\footnote{Email: mssp@uohyd.ernet.in}.

{\scriptsize {
$^a$ {\it Theory Division, Saha Institute of Nuclear Physics, 1/AF Bidhannagar, Calcutta 700064, India}.\\
$^b$ {\it School of Physics, University of Hyderabad, Hyderabad 500046, India}.\\
$^c$ {\it School of Mathematical Sciences, UCD, Belfield, Dublin 4, Ireland}.\\
$^d$ {\it Department of Theoretical Physics, Indian Association for the Cultivation of Science, Calcutta  700032, India}.\\ }}

\end{center}
\vskip.5cm

\begin{abstract}
The AdS/CFT conjecture offers the possibility of a quantum description for a black hole in terms of a CFT. This has led
to the study of general $AdS^3$ type black holes with a view to constructing an explicit toy quantum black hole model. Such a  CFT description 
would be characterized by its central charge and the dimensions of its primary fields. Recently the expression for the central charges $(C_L, C_R)$ of the CFT dual to the warped $AdS^3$ have been determined using asymptotic symmetry arguments. The central charges depend, as expected, on the warping factor. We show that topological arguments, used by Witten to constrain central charges for the BTZ black hole, can be generalized to deal with the warped $AdS^3$ case. Topology constrains the warped factor to be rational numbers while quasinormal modes are conjectured to give the dimensions of primary fields. We find that in the limit when warping is large or when it takes special rational values the system tends to Witten's conjectured unique CFT's with central charges that are multiples of 24.
\end{abstract}
\vspace*{.3cm}
\begin{center}
March 2010
\end{center}
\vspace*{1.0cm}

The possibility of constructing a complete quantum theory of gravity for the 3D BTZ was attempted by Witten \cite{witten}.The approach used the AdS/CFT
correspondence and constructed the quantum theory of gravity as a CFT which had holomorphic factorization and central charges given by the pair ($24k_1, 24 k_2$) where $k_1, k_2 \in Z$. For $k_1 = k_2 = 1$, the corresponding CFT was conjectured to be unique.Subsequent work suggested that a  pure 3D quantum gravity theory might not be physically sensible \cite{maloney}.

Recently a generalization of the BTZ black hole, the warped $AdS^3$ black hole, arose in the context of topologically massive gravity \cite{wAds3, wAds31, wAds32,wAds33}.  Using analogies suggested by the BTZ black hole \cite{btz,btz1}, the central charges $C_L, C_R$ of the holomorphic and anti-holomorphic dual CFT's of the warped $AdS^3$ black  holes were first conjectured in \cite{strom} and later these  conjectured forms of $C_L$ and $C_R$ shown to
follow from a study of the asymptotic symmetry of the warped $AdS^3$ black hole\cite{steph1,steph2,mbbc,omro}. The central charges were found to respect the gravitational anomaly constraint.

In view of these results we study if there are topological restriction on the central charge for the warped $AdS^3$ black hole, following the analysis of \cite{strom}. We note that the warped $AdS_3$ system has vector type excitations present and hence is not a pure gravity theory. Following the analysis of Witten \cite{witten} we find that topology restricts warping to rational numbers and that surprisingly for large warping the associated CFT tends to a pure gravity theory. This follows from the observation that in this limit the central charges for the associated CFT's are integral multiples of 24. Such CFT's, were conjectured by Witten \cite{witten} to be unique to represent pure gravity theories. We next observe that, following the treatment in \cite{danny1,danny2}, that the conformal weights of the associated CFT's for the warped $AdS^3$ can also be  determined from  their quasinormal modes\cite{us,oh,wen,chen}. Thus the associated dual CFT's are completely determined in terms of the warping factor and the quasinormal modes of the warped $AdS^3$ black hole.

The action for topologically massive gravity in three dimensions in terms of two Chern-Simons gauge theories is given by
\begin{equation}
 S_{EH}=\frac{K_L+K_R}{2}(I_{CS}(A_L)-I_{CS}(A_R))\label{eh}
\end{equation}
and the gravitational Chern-Simons action has the form
\begin{equation}
 S_{GCS}=\frac{K_L-K_R}{2}(I_{CS}(A_L)+I_{CS}(A_R)),\label{gcs},
\end{equation}
where $I_{CS}$ is the Chern-Simon action given by 
\begin{equation}
 I_{CS}(A)=\frac{1}{4\pi}\int d^3 x ~tr \left( A\wedge dA+\frac{2}{3}A\wedge A\wedge A\right).\label{cs}
\end{equation}
The allowed values of the coupling constants $K_L, K_R$ get restrictions, if the gauge group is $SO(2,1)\times SO(2,1)$. The BTZ black hole has the symmetry under $SL(2,R)\times SL(2,R)$. Witten's idea was to think of $SL(2,R)$ as a double cover of the $SO(2,1)$ \cite{witten}.From the fact that $SO(2,1)$ has a compact $SO(2)$ subgroup in it, topological considerations restrict $K_L, K_R$ values. We would like to investigate if Witten's restrictions are relevant for the warped $AdS^3$ black hole.

Consider the the action
\begin{equation}
 S=\frac{1}{16\pi G}\int d^3 x\sqrt{-g}(R+\frac{2}{l^2}) +\frac{l}{96\pi G\nu}\int d^3 x \sqrt{-g}\epsilon^{\mu\nu\lambda}\Gamma_{\mu\sigma}^\rho \left(\partial_\lambda\Gamma_{\rho\lambda}^\sigma +\frac{2}{3}\Gamma_{\nu\beta}^\sigma\Gamma_{\lambda\rho}^\beta\right)\label{wtmg}
\end{equation}
describing topologically massive gravity in three dimension.
Here the warping factor $\nu=\mu l/3$ where $\mu$ is the mass and $\epsilon^{\mu\nu\lambda}=+1/\sqrt{-g}$. The second term in the above is the gravitational Chern-Simon term -$I_{GCS}$.
The warped black hole with $U(1)\times SL(2,R)$ symmetry is a classical solution to the above action which goes to warped $AdS^3$ asymptotically \cite{strom}. The warping was done by constructing a fibration of $AdS^3$ as $R\times SL(2,R)$. The resulting spacetime still has a constant negative curvature as $AdS^3$ but have a very different metric.

In order to get topological constraints we follow Witten\cite{witten} and replace $SL(2,R)$ by $SO(2,1)$ and rewrite the above action in Eqn.(\ref{wtmg}) as a $SO(2,1)$ gauge theory. We find that the action can be written as the sum of two Chern Simon terms of opposite chirality. This immediately allows us to use the toplogical ideas of Witten. We introduce $A_L=\omega-{}^*e/l$, 
$A_R=\omega+{}^*e/l$ where ${}^*e=\epsilon^{abc}e_c$ is a two index object in $SO(2,1)$ and $\omega$ is the spin connection. Using these $A_L$ and $A_R$ in Eqns.(\ref{eh},\ref{gcs}), we find
\begin{equation}
 \frac{1}{\pi l}\int d^3 x\sqrt{-g}(R+\frac{2}{l^2}) =-\frac{1}{\pi l}\int d^3 x ~tr {}^*e (d\omega \wedge 
\omega)-\frac{l}{3\pi l^3}\int d^3 x ~tr ({}^*e\wedge{}^*e\wedge{}^*e).
\end{equation}
The gravitational Chern-Simon action $I_{GCS}$ can be written in terms of spin connection $\omega$ alone as $\int tr~(1/2\pi)(\omega\wedge d\omega+(1/3\pi)\omega\wedge\omega\wedge\omega)$.

Thus the action for warped, topologically massive gravity given is Eqn.(\ref{wtmg}) can be expressed in terms of the Chern-Simon action for the gauge fields $A_L$ and $A_R$ as
\begin{equation}
 S=S_{EH}+S_{GCS}\label{wtmg1}
\end{equation}
where $S_{EH}$ and $S_{GCS}$ are given in Eqns.(\ref{eh},\ref{gcs}).
Comparing the coefficients of the above action with that in Eqn. (\ref{wtmg}), we get
\begin{eqnarray}
 \frac{K_L+K_R}{2\pi l}&=&\frac{1}{16\pi G}\label{kplus}\\
\frac{K_L-K_R}{4\pi l}&=&\frac{1}{94 \pi G\nu}\label{kminus},
\end{eqnarray}
which gives
\begin{equation}
 \frac{l}{G\nu}=24(K_L-K_R).
\end{equation}

Exploiting the fact that the action in Eqn.(\ref{wtmg1}) is expressed in terms of the Chern-Simons theory, Witten used the topological arguments to conclude that 
\begin{equation}
K_L \in Z/2,~~K_L-K_R\in Z.\label{cond}
\end{equation}
Here the fact that $SL(2,R)$ is double cover of $SO(2,1)$ is used. The constraints in Eqn.(\ref{cond}) along with
Eqns.(\ref{kplus},\ref{kminus}) imply
\begin{eqnarray}
 \frac{l}{G\nu}&=&24 n\label{n}\\
\frac{l}{8G}&=&\frac{m}{2}\label{n2},~~~~~n,m\in Z.
\end{eqnarray}
From the above we also get $\nu=m/6 n$.Thus topology forces the warping factor to be a rational number.

Warped $AdS^3$ black hole which is asymptotic to warped $AdS^3$ is free of naked closed time like curves and other pathologies arise as solution to the warped, topologically massive gravity with $\nu^2>1$ \cite{clement1,clement2}. Note that although the generic structure of warped $AdS^3$ black hole is very different and have reduced symmetry compared to the unwarped case, the topological constraints of Witten which are independent of metric are still relevant.

We now turn to the explicit forms of $C_L$ and $C_R$ obtained in \cite{strom} which are
\begin{eqnarray}
 C_L&=&\frac{4\nu l}{G(\nu^2+3)}\label{cl}\\
C_R&=&\frac{(5\nu^2+3)l}{G\nu(\nu^2+3)}.\label{cr}
\end{eqnarray}
These results were confirmed using different approaches in \cite{mbbc, omro}.
Furthermore, these central charges also obey
\begin{equation}
 C_L-C_R=-\frac{l}{G\nu},\label{cchargedif}
\end{equation}
which matched the diffeomorphism anomaly\cite{pkfl}.

Using Eqns.(\ref{n}) in Eqn.(\ref{cchargedif}) we find 
\begin{equation}
C_L-C_R=-24n, ~n\in Z
\end{equation}
which agrees with the requirement of modular invariance that the difference between the left and right moving ground state energies must be an integer. Substituting Eqn.(\ref{n}) in Eqn.(\ref{cchargedif}) and using Eqns.(\ref{cl},\ref{cr}), we find
\begin{eqnarray}
 C_L&=&24 n \left(\frac{4 \nu^2}{\nu^2+3}\right),\label{cl1}\\
C_R&=&24n\left(1+\frac{4\nu^2}{\nu^2+3}\right).\label{cr1}
\end{eqnarray}
There are several observations one can make about Eqns. (\ref{cl1}) and (\ref{cr1}). Note that for the usual BTZ black hole in 3d gravity, Witten argued that the central charges of the left and right moving sectors must be of the form $(24k_1, 24k_2)$ with $k_1,k_2$ integers. The CFT with a  central charge $24k$ with $k=1$ was constructed by Frenkel, Lepowsky and Meurman \cite{flm} and conjectured to be unique. Witten suggested that this unique CFT with
an associated Monster Group symmetry was the quantum theory of the BTZ black hole. He further conjectured that such unique CFT's should also exist for $k>1$ \cite{witten}. We note that the warped $AdS_3$ black hole central charges tend to multiples of 24 in the limit of large warping or for special values. This means that in these cases the black hole system tends to a conjectured pure gravity theory with a unique CFT description. In these cases the central charges are of the form $(24k_1, 24k_2)$,which is precisely the case considered by Witten. The special values for the warping factor for which this happens are:
\be \label{warprestrict}
\nu^2 = \frac{3k_1}{4k_2 - 5k_1}.
\ee 
We also note that when $\nu^2>>1$, for a fixed $n$,
\begin{equation}
 (C_L,C_R)\to (24n_1, 24n_2),~~n_1=4n,~n_2=5n.
\end{equation}
The large $n_1, n_2$ limit which we need to make contact with these values for central charges is expected to describe classical black hole. This is so as it is only for large `quantum charges'that  the quantum system is expected to have a classical description. In this limit all the thermodynamics quantities, $S, T_L, T_R$ become large and classical. In general, the central charges do not have the $(24k_1, 24k_2)$ structure, for warped topologically massive gravity. However in these cases the central charges are still determined to be rational multiples of 24 with associated primary fields, whose dimension can be read off from the quasinormal modes that have been determined. It would be of interest to exploit this quantum information
to carry out dynamical calculations.

\bigskip
\noindent
{\bf Acknowledgments}

One of us (SS) thank School of Physics, University of Hyderabad, where a part of this work was done while visiting as Jawaharlal Nehru Chair Professor.


\begin{thebibliography}{99}
\bibitem{witten} E. Witten, `Three-Dimensional Gravity Revisited', arXiv:0706.3359[hep-th].
\bibitem{maloney} A. Maloney and E. Witten, JHEP {\bf 1002}, 029 (2010). 
\bibitem{wAds3}S. Deser, R. Jackiw and S. Templeton, `Topologically Massive Gauge Theories',
Annals. Phys. {\bf 140}, 372 (1982).
\bibitem{wAds31} S. Deser, R. Jackiw and S. Templeton, `Three-Dimensional Massive Gauge Theories', Phys. Rev. Lett. {\bf 48}, 975 (1982).
\bibitem{wAds32} W. Li, W. Song and A. Strominger,`Chiral Gravity in Three Dimensions', JHEP {\bf 0804}, 082 (2008).
\bibitem{wAds33} S. Carlip, S. Deser, A. Waldron and D.K. Wise, `Topologically Massive AdS Gravity', Phys. Lett. {\bf B 666}, 272 (2008).
\bibitem{btz} M. Banados, C. Teitelboim and J. Zanelli, `Black hole in three-dimensional spacetime', Phys. Rev. Lett. {\bf 69}, 1849 (1992).
\bibitem{btz1} M.~Banados, M.~Henneaux, C.~Teitelboim and J.~Zanelli, `
Geometry of the 2+1 black hole', Phys. Rev.{\bf  D 48}, 1506 (1993).
\bibitem{strom} D. Anninos, W. Li, M. Padi, W. Song and A. Strominger, 
`Warped $AdS_3$ Black Holes', JHEP {\bf 0903} (2009) 130, arXiv:0807.3040 [hep-th]
\bibitem{steph1} G. Compere and S. Detournay, Class. Quant. Grav. {\bf 26},012001 (2009).
\bibitem{steph2} G. Compere and S. Detournay, JHEP {\bf 0908}, 092 (2009). 
\bibitem{mbbc} M. Blagojevic and B. Cvetkovic, `Asymptotic structure of topologically massive gravity in spacelike stretched AdS sector', JHEP {\bf 0909} (2009)006, arXiv:0907.0950[hep-th].
\bibitem{omro} O. Miskovic and R. Olea,`Background-independent charges in Topologically Massive Gravity',arXiv:0909.2275[hep-th].
\bibitem{danny1} D. Birmingham, I. Sachs, S. N. Solodukhin, `Conformal Field Theory Interpretation of Black Hole Quasi-normal Modes',Phys. Rev. Lett. {\bf 88}, 151301 (2002). 
\bibitem{danny2} D. Birmingham, I. Sachs, S. N. Solodukhin, `Relaxation in Conformal Field Theory, Hawking-Page Transition, and Quasinormal/Normal Modes',Phys. Rev. {\bf D 67}, 104026 (2003). 
\bibitem{us} K. S. Gupta, E. Harikumar, S. Sen and M. Sivakumar, `Geometric Finiteness, Holography and Quasi Normal Modes for the Warped ${\rm AdS}_3$ Black Hole', arXiv:0912.3584 [hep-th].
\bibitem{oh} J. J. Oh and  W. Kim, `Absorption Cross Section in Warped AdS$_3$ Black Hole',   JHEP {\bf 0901} (2009) 067.
\bibitem{wen} H.-C. Kao and W.-Y. Wen, `Absorption cross section in warped AdS$_3$ black hole revisited', JHEP {\bf 0909} (2009) 102
\bibitem {chen} B. Chen and Z.-B. Xu, `Quasi-normal modes of warped black holes and warped AdS/CFT correspondence', JHEP {\bf 0911} (2009) 091.
\bibitem{clement1} A. Bouchareb and G. Clement, Class. Quant. Grav. {\bf 24} (2007) 5581.
\bibitem{clement2} K. A. Moussa, G. Clement, H. Guennoune and C. Leygnac, Phys. Rev. {\bf D78} (2008) 064065.
\bibitem{dbss} D. Birmingham and S. Sen, ott Time Machines, `BTZ Black Hole Formation, and Choptuik Scaling', Phys. Rev.Lett. {\bf 84} (2000) 1074, hep-th/9908150.
\bibitem{pkfl} P. Kraus and F. Larsen, `Holographic gravitational anomalies', JHEP {\bf 0601} (2006) 022, hep-th/0508218.
\bibitem{flm} I. B. Frenkel, J. Lepowski and A. Meurman, Proc. Natl. Acad. Sci. USA {\bf 81}, 3256 (1984).

\end{thebibliography}
\end{document}